\documentclass{article}
\usepackage[a4paper]{geometry}

\usepackage[utf8]{inputenc}
\usepackage{graphicx}
\usepackage{comment}
\usepackage{xcolor}
\usepackage{multirow} 
\usepackage{hyperref}
\usepackage{pdfpages}
\usepackage{caption, subcaption}
\usepackage{amsmath, amssymb}
\usepackage{cite}
\usepackage{enumitem}

\usepackage{versions}
\excludeversion{paper}
\includeversion{extended}

\begin{document}

\begin{paper}
\title{Evaluating Social Acceptance of eXtended Reality (XR) Agent Technology: A User Study}
\end{paper}

\begin{extended}
\title{Evaluating Social Acceptance of eXtended Reality (XR) Agent Technology: A User Study (Extended Version)}
\end{extended}

\author{
Megha Quamara \\
Department of Informatics, King’s College London, London, UK \\
\texttt{megha.quamara@kcl.ac.uk}
\and
Viktor Schmuck \\
Department of Engineering, King’s College London, London, UK \\
\texttt{viktor.schmuck@kcl.ac.uk}
\and
Cristina Iani \\
Department of Surgery, Medicine, Dentistry and Morphological Sciences, \\
University of Modena and Reggio Emilia, Reggio Emilia, Italy \\
\texttt{cristina.iani@unimore.it}
\and
Axel Primavesi \\
Deutsche Welle, Berlin, Germany \\
\texttt{axel.primavesi@dw.com}
\and
Alexander Plaum \\
Deutsche Welle, Bonn, Germany \\
\texttt{alexander.plaum@dw.com}
\and
Luca Vigan\`o \\
Department of Informatics, King’s College London, London, UK \\
\texttt{luca.vigano@kcl.ac.uk}
}

\date{}

\maketitle

\begin{abstract}
In this paper, we present the findings of a user study that evaluated the social acceptance of eXtended Reality (XR) agent technology, focusing on a remotely accessible, web-based XR training system developed for journalists. This system involves user interaction with a virtual avatar, enabled by a modular toolkit. The interactions are designed to provide tailored training for journalists in digital-remote settings, especially for sensitive or dangerous scenarios, without requiring specialized end-user equipment like headsets. Our research adapts and extends the Almere model, representing social acceptance through existing attributes such as perceived ease of use and perceived usefulness, along with added ones like dependability and security in the user-agent interaction. The XR agent was tested through a controlled experiment in a real-world setting, with data collected on users' perceptions. Our findings, based on quantitative and qualitative measurements involving questionnaires, contribute to the understanding of user perceptions and acceptance of XR agent solutions within a specific social context, while also identifying areas for the improvement of XR systems.
\end{abstract}


\section{Introduction}
\label{Sec:Introduction}

Extended Reality, or XR, brings together technologies like Virtual Reality (VR), Augmented Reality (AR), and Mixed Reality (MR), which combine digital elements with the physical (or~real) world to deliver interactive and immersive user experiences~\cite{stanney2021extended}. XR-based agent solutions take such experiences a step further by utilizing Artificial Intelligence (AI)-powered virtual assistants or conversational agents within these immersive environments~\cite{hirzle2023xr}. These synthetic agents can mimic real-life interactions to provide personalized experiences that enhance learning and decision-making. These offer valuable tools for applications in fields like education~\cite{petersen2021pedagogical} and healthcare~\cite{deighan2023social}, enabling users to interact with technology in more natural, flexible ways. 

XR systems can also be highly beneficial in fields like journalism training, offering immersive, hands-on experiences that surpass what traditional methods like books or even on-the-job training can provide. Journalists can practice scenarios like interviewing under pressure or reporting in conflict zones, without real-world risks. However, existing XR training solutions present challenges~\cite{mabrook2019virtual}; they require expensive equipment, high costs per trainee, and a physical location for training. Many also exhibit poor human-computer interaction and provide limited personalization based on users' needs. The use of conversational agents, however, can help overcome these limitations by offering more cost-effective, flexible, and engaging training experiences while enhancing user interaction.

If XR-based conversational agents are to achieve wide-scale adoption across such applications, it is essential to understand how different user groups accept or reject this technology, as skepticism can hinder its integration into various social contexts~\cite{tabourdeau2020user}. How will users respond to interacting with virtual avatars? To what extent do differences in the agent’s conversational and social abilities, such as responsiveness, expressiveness, or realism, affect users' attitudes and willingness to engage with the system? How do users perceive the security, privacy, and trustworthiness of such agents in varying deployment contexts? Understanding these aspects is essential not only to assess the feasibility of deploying such systems at scale but also to guide the development of more effective, user-centered tools that users trust and are motivated to use, ultimately contributing to their acceptance.

In this paper, we report on the findings of a user study that we conducted to evaluate the social acceptance of XR agent technology, focusing on a remotely accessible, web-based XR training system developed for journalists. This system involves interaction of the users with a virtual avatar, which is enabled by a modular toolkit. The interactions are designed to deliver tailored training for journalists in digital and remote environments, particularly in sensitive and high-risk situations, without the need for specialized equipment, such as headsets. We employed a mixed-methods approach, combining quantitative questionnaire analysis and qualitative interviews, in order to get a deeper understanding of the attitudes and perception of potential future users. Our research builds on the \emph{Almere model}~\cite{heerink2010assessing}, which we adapt and extend through a customized questionnaire to better understand the perceptions of the participants interacting with the training system in a controlled, real-world setting. The questionnaire includes questions regarding existing attributes, such as perceived usefulness and ease of use, adapted to the context of our case study, along with additional questions related to security, privacy, and trust. The system has been deployed as a prototype, and while it was under development, the study that we carried out allowed us to identify actionable considerations for its implementation, as the prototype is under continuous improvement. This study aims to inform standard practices for evaluating XR-based solutions by offering insights into their social acceptance in learning and training contexts, as well as in service-oriented settings such as post offices, information points, and customer service. Our findings, based on quantitative and qualitative data from the questionnaires, support this aim by enhancing the understanding of user perceptions and acceptance of XR agent solutions, while also identifying areas for improvement in system design and implementation.

We proceed as follows. Section~\ref{Sec:SocialAcceptance} presents the concept of social acceptance. Section~\ref{Sec:RelatedWork} reviews related work. Section~\ref{Sec:OurStudy} details our study, including methodology. Section~\ref{Sec:Findings} presents our findings. Section~\ref{Sec:Conclusion&FutureWork} concludes and discusses future work. \begin{extended} The Appendix~\ref{Sec:UserConsentForm} contains the user consent form which the participants signed before taking part in this study. \end{extended}

\section{Social Acceptance}
\label{Sec:SocialAcceptance}

Acceptance of technology has been understood in various ways in the literature. It is commonly associated with the intention to use (or interact with) a specific technology following initial use~\cite{venkatesh2003user,villena2023extended}. Acceptance can be viewed from an individual user's perspective (i.e., \emph{user acceptance})~\cite{distler2018acceptability} or from a broader social perspective (i.e., \emph{social acceptance})~\cite{taebi2017bridging}, which refers to the extent to which a community accepts, or at least tolerates, a technology. Thus, acceptance depends not only on the technology itself but also on the users who interact with it and relate to it.

In the case of XR, we posit that social acceptance depends on the system's ability to provide users with an experience that
\begin{itemize}[noitemsep]
    \item is satisfactory in that the system provides interactions and responses that meet the goals and needs of the user, while also being compliant with the social context in which it is used (meaning that the system adheres to the requirements appropriate to that context, thereby ensuring that it can be employed across diverse settings, without making users feel uncomfortable or out of place), 
    \item is transparent in that the data, system, and models should be presented with clear explanations of how they work and make decisions, and should be easily accessible so that users can interact with the system in a user-friendly manner,
    \item is safe and benefits all human users, regardless of age, gender, race, language, or appearance,
    \item is secure in that it guarantees communication security and data privacy, and
    \item is explainable and trustworthy in that it is reliable and ensures trust in the system's responses.
\end{itemize}

These features have also been discussed in prior work on XR agents and human-agent interaction. We elaborate on these perspectives in the related work section, where we review applications of XR agent technology in training and approaches to measuring social acceptance.

\section{Related Work}
\label{Sec:RelatedWork}

\subsection{Applications of XR Agent Technology in Training}
\label{Sec:ApplicationsOfXRAgentTechnologyInTraining}

XR technologies are increasingly being applied in education and professional training to enhance engagement, knowledge acquisition, and knowledge transfer~\cite{gan2023researching,fidan2019integrating,mena2023teachers}. They have demonstrated strong potential in safety training within high-risk sectors, such as construction and oil and gas~\cite{scorgie2024virtual}. Furthermore, studies have shown the effectiveness of XR technologies in supporting skill development in various domains. These include soft skills and communication~\cite{van2020serious}, clinical reasoning in healthcare education~\cite{sim2022virtual}, patient education~\cite{curran2024use}, emergency preparedness~\cite{khanal2022virtual}, and decision-making and leadership in business training~\cite{alcaniz2018virtual}. 

Building upon the advantages of intelligent virtual agents and AI-powered characters, which represent an evolution of chatbots, XR agent technology has shown potential in facilitating learning~\cite{fink2024ai}. Although chatbots interact with users via natural language, they are mostly text-based. In contrast, AI-powered avatars embody a human. In addition, recent technological advances have increased their ability to interact more dynamically with users, thus providing engaging, adaptive, and personalized learning experiences~\cite{dai2024effects}.

In journalism, a field that is not yet widely explored in this context, XR agent applications may enable simulations of hazardous environments and interactions, helping journalists train to handle unpredictable challenges, respond to emergencies, and manage conflict and stress~\cite{hoak2025virtually}, while exploiting the benefits of AI-based avatars and their human-like interactive abilities to increase engagement, and consequently, learning effectiveness~\cite{dai2024effects}. These benefits may be further enhanced by the development of web-based XR applications, which can reduce costs, facilitate remote learning, thereby increasing application contexts, alleviate discomfort associated with headset use, and enhance interactivity~\cite{shaikh2022data}.

\subsection{Measuring Social Acceptance of XR Agents}
\label{Sec:MeasuringSocialAcceptanceOfXRAgents}

The immersive and impactful learning experiences enabled by XR agent technology can only be effective if users are willing to engage with it. Therefore, the design and implementation of such systems must account for users' social acceptance, which is a key prerequisite for technology adoption. A lack of social acceptance can become a barrier, discouraging some users from using the technology and preventing them from accessing the services and information it provides. In recent years, there has been a growing body of research aimed at identifying the key features that can affect user perception and enhance users' social acceptance of XR technology~\cite{tabourdeau2020user}. These studies have primarily focused on examining user acceptance of XR and its constituent technologies, such as VR~\cite{sagnier2020user, jang2021augmented} and AR~\cite{jang2021augmented}, across various application contexts. The research has often employed frameworks like the \emph{Technology Acceptance Model (TAM)}~\cite{lee2003technology, davis1985technology} and the \emph{Unified Theory of Acceptance and Use of Technology (UTAUT)}~\cite{williams2015unified,taherdoost2018review}.

In the context of XR technologies, studies have extended TAM by integrating additional variables such as perceived enjoyment~\cite{manis2019virtual}, personal and situational factors~\cite{chung2015tourists}, and specific XR characteristics~\cite{makransky2018structural}. Several factors influencing the adoption of XR technologies have been identified, including sense of presence~\cite{sagnier2020user} and perceived ease of use~\cite{manis2019virtual}. However, it should be noted that research on the social acceptance of XR agent solutions remains limited, with most existing studies focusing on features of human–virtual avatar interaction~\cite{kyrlitsias2022social}. This is particularly true with the use of XR in journalists' training. In line with our discussion in Section~\ref{Sec:SocialAcceptance}, the assessment of social acceptance for XR agents should consider the system's ability to deliver an experience that meets user goals and needs, aligns with the social context of use, and is perceived as transparent, safe, secure, explainable, and trustworthy. Relevant to this latter point, even though trust is recognized as one of the factors affecting social acceptance, few studies assessed it in the context of XR agent technologies~\cite{rheu2021systematic}. 

\section{Our Study}
\label{Sec:OurStudy}

Our study evaluates the social acceptance of XR agent technology, focusing on user experiences with an XR-based and a web-based training system developed for journalists. Conducted in the context of the media organization Deutsche Welle, the study focuses on interactions of journalists with a remote XR training solution, named \emph{Guardia}. We employed a mixed-methods approach, combining questionnaire data with interviews, to investigate the perceptions of the participants. In the following, we detail the system \emph{Guardia}, describe our study’s methodology and procedure, outline the questionnaire development, and describe the data analysis method we used.

\subsection{Artificial Security Agent}
\label{Subsec:ArtificialSecurityAgentGuardia}

Deutsche Welle (DW) is Germany’s international media organization with a wide network of correspondents and other staff working worldwide. Many of these individuals work in high-risk or conflict zones and crises, making security training for them a critical priority. The organization has a dedicated team focused on protecting staff in such challenging environments and ensuring their safety, with staff training being one of the team's key responsibilities. The training process is time-consuming, costly, and labor-intensive, requiring significant resources to ensure that all individuals are adequately prepared for the risks they may face (e.g., dangers evoked by hostile groups who disapprove of journalists, or sensitive cultural and diversity issues faced in interview situations).

\begin{figure*}[t]
    \begin{subfigure}[b]{0.45\textwidth}
        \centering
        \includegraphics[height=4cm, width=4cm]{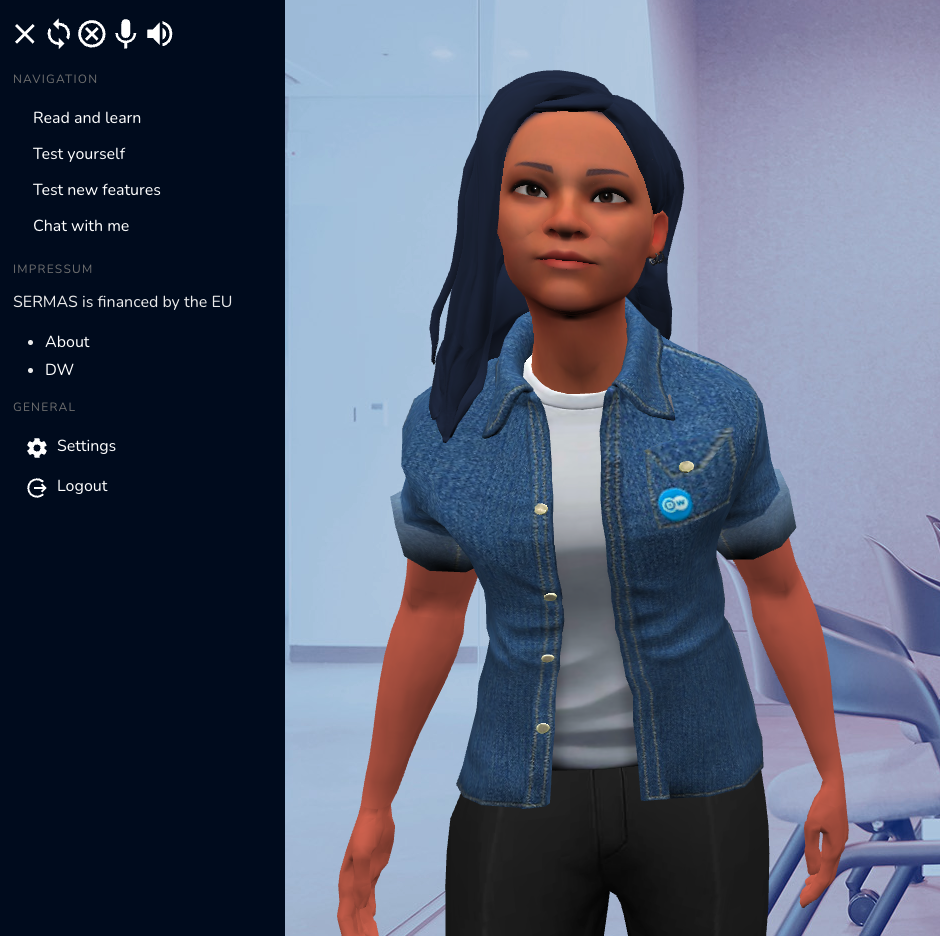}
        \caption{Main Menu}
        \label{Fig:MainMenu}
    \end{subfigure}
    \hfill
    \begin{subfigure}[b]{0.45\textwidth}
        \centering
        \includegraphics[height=4cm, width=6cm]{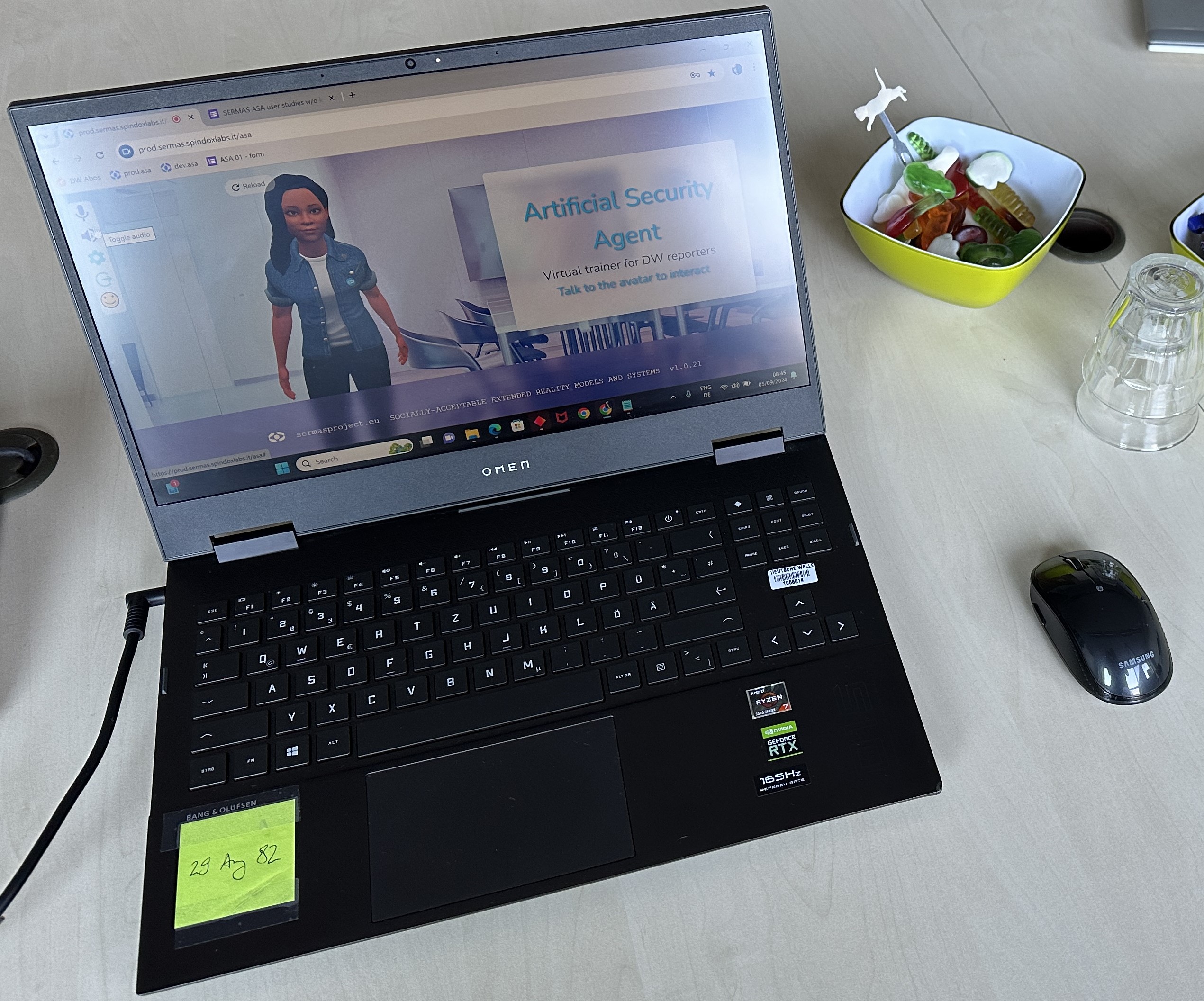}
        \caption{User Study Setup}
        \label{Fig:UserStudySetup}
    \end{subfigure}
    \begin{subfigure}[b]{0.45\textwidth}
        \centering
        \includegraphics[height=4.0cm, width=7cm]{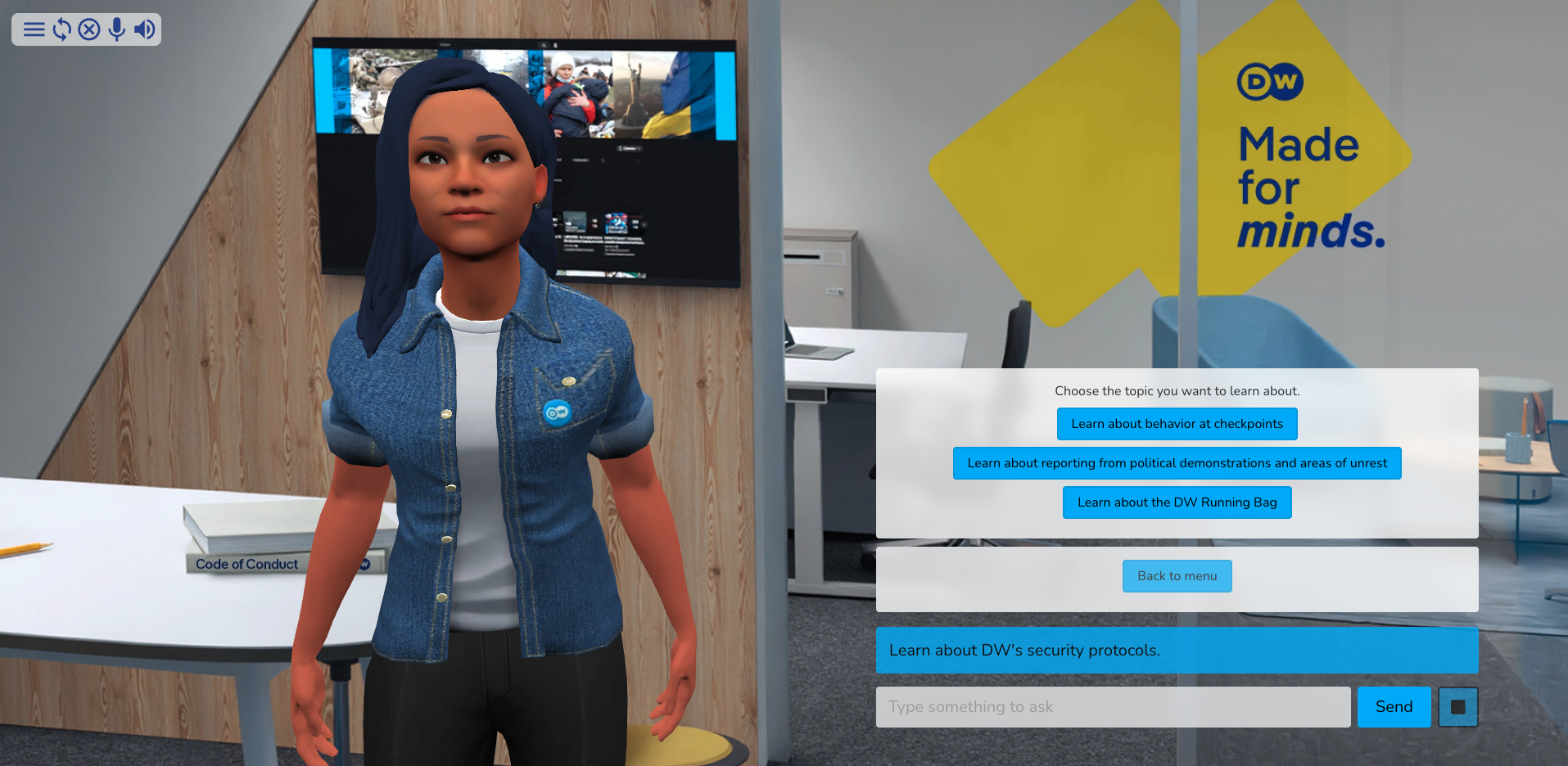}
        \caption{Lesson Overview}
        \label{Fig:LessonOverview}
    \end{subfigure}
    \hfill
    \begin{subfigure}[b]{0.45\textwidth}
        \centering
        \includegraphics[height=4.0cm, width=5cm]{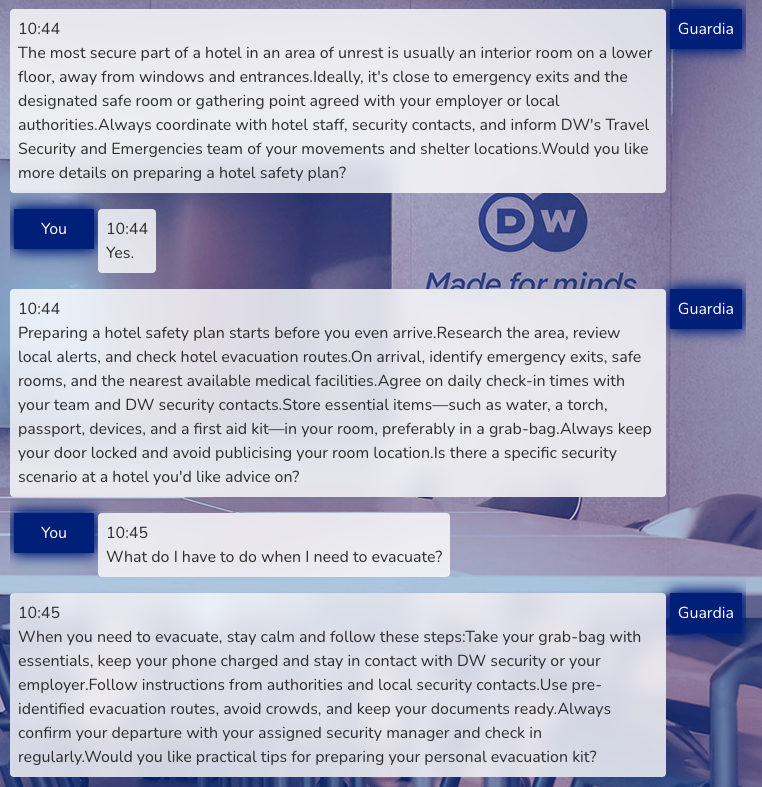}
        \caption{Free Chat Interaction}
        \label{Fig:FreeChatInteraction}
    \end{subfigure}
    \caption{Guardia User Interface.}
    \label{Fig:GuardiaUserInterface}
\end{figure*}

The idea behind the training, hosted by the security agent as a virtual co-trainer, is to support the security team in providing the journalists with the basic information on various topics, such as how to behave at checkpoints (police, military, or militia), how to report from political demonstrations or areas of unrest, and what equipment is needed in emergencies. This involves an XR and web-based training solution customized for journalists, developed within the context of the SERMAS project.\footnote{\emph{SERMAS (Socially-acceptable Extended Reality Models and Systems)} is a 3-year Horizon Europe research and innovation project that started on October 1, 2022. The project aims to improve human-machine interaction by developing new models and systems for XR that offer higher levels of interaction and greater context awareness, laying the foundations for next-generation XR systems focused on socially-acceptable interactions. \\See \url{https://sermasproject.eu/}.} The solution combines a platform-agnostic virtual avatar with an AI-driven chatbot powered by a Large Language Model (LLM), enabling personalized and practical training across a range of journalistic scenarios (see Figure~\ref{Fig:MainMenu}). The main architectural components underlying this training solution are detailed in~\cite{quamara2024towards}. The agent named \emph{Guardia} is not intended to replace real-life training by human instructors, nor is it capable of doing so. Instead, it offers participants an innovative and engaging way to learn the basics remotely, using realistic learning scenarios without the need for specialized end-user equipment like headsets.

\subsection{Methodology and Procedure}
\label{Subsec:MethodologyAndProcedure}

To evaluate the social acceptance by journalists of the virtual training agent \emph{Guardia}, a user study was organized by members of the DW unit for research and cooperation projects, which took place in September 2024 at DW's premises in Bonn and Berlin. The study was approved by the Ethical Committee of the University of Modena and Reggio Emilia (protocol n. 2024-UNMRCLE-0198640), which is another 
partner of the SERMAS project. It involved 25 participants, all of whom were DW's staff members, and were selected by the members of the aforementioned unit. They work either in the editorial department as journalists, in the production department as producers, or at the DW academy as trainers. They have experience working in potentially dangerous environments and have received security training in the past, or may do so in the future. In terms of gender distribution, 48\% of the participants identified themselves as female, 44\% as male, 4\% as non-binary, and 4\% preferred not to disclose their gender identity. In terms of experience with virtual agents, 18 participants had interacted with them a few times, 3 were aware of them but had never interacted with them, 3 had worked with/used them, and 1 had no exposure to them. In terms of age distribution, the majority of participants (44\%) were in the 45–54 age group, indicating that middle-aged individuals formed the largest segment of the sample. This was followed by 24\% in the 35–44 age range and 20\% in the 25–34 group, suggesting a reasonable representation of younger adults as well. Only 12\% were in the 55–64 age range, indicating lower participation from older adults.

Each participant attended a single session with a member of the aforementioned unit in a meeting room on the premises. During the session, the facilitator explained the experiment procedure and introduced the participants to an early prototype of the virtual co-trainer \emph{Guardia}. Participants used a standard laptop with a mouse connected to it (see Figure~\ref{Fig:UserStudySetup}). They were informed that the system would access the laptop’s microphone and webcam during the training, but no data would be stored. 
\begin{paper}
The participants, thus, could interact with \emph{Guardia} using voice, keyboard, and mouse. Before participating in the training session, all participants provided their consent by signing a consent form provided in~\cite{quamara25}.
\end{paper}
\begin{extended}
The participants, thus, could interact with \emph{Guardia} using voice, keyboard, and mouse. Before participating in the training session, all participants provided their consent by signing a consent form (see Appendix~\ref{Sec:UserConsentForm}).
\end{extended}

After a short introduction, the participants were asked to complete the following tasks: (1)~three lessons presented by \emph{Guardia}, covering the following topics (see Figure~\ref{Fig:LessonOverview}): behavior at checkpoints, reporting from political demonstrations or areas of unrest, and the contents of the organization's running bag (standard equipment for emergencies); (2) three quizzes on these topics; and finally, (3) at least one chat interaction with \emph{Guardia} regarding security issues (see Figure~\ref{Fig:FreeChatInteraction}). During the training, the facilitator was responsible for monitoring and recording interventions, the length of each part of the training, the duration and mode of the entire interaction, and whether the participants completed all the lessons and quizzes with or without assistance.

After the training, an online questionnaire and a short interview were conducted, with responses (transcriptions) stored anonymously. The questionnaire, which served as the data collection instrument, consisted of 51 items on a 5-point Likert scale ranging from 1 to 5 (totally disagree --- disagree --- don’t know --- agree --- totally agree). The items (questions) were presented to the participants in a random order and were alternated between positive and negative phrasing to avoid acquiescence bias and to ensure that participants engage more thoughtfully with the content of each item~\cite{friborg2006likert}. 

Upon completing the questionnaire, participants were asked 10 open-ended questions aimed at gathering additional information about their experience with \emph{Guardia} and the training to help clarify the results of the questionnaire.

\subsection{Questionnaire Development}
\label{Subsec:QuestionnaireDevelopment}

We chose to adapt and extend the \emph{Almere questionnaire}~\cite{heerink2010assessing} for our research because it is based on the \emph{UTAUT}~\cite{williams2015unified}, one of the most established frameworks guiding the assessment of social acceptance. The Almere model includes dimensions such as Perceived Ease of Use, Perceived Usefulness, and Social Influence, which align closely with the features and capabilities XR technologies must offer to deliver experiences as discussed in Section~\ref{Sec:SocialAcceptance}. Additionally, the Almere questionnaire has demonstrated strong reliability and validity~\cite{heerink2010assessing}, making it a robust tool for evaluating social acceptance in our context of a virtual security agent. We added Security, Informational Privacy, Trust, and Dependability as new attributes to the existing model, as these factors have been found to be influential in several other studies (see e.g.,~\cite{katins2024assessing,harborth2021investigating}). 

In Table~\ref{Tab:RepresentativeAttributes}, we define the attributes that we identified as potential direct determinants or representatives of social acceptance. This is based on their conceptual alignment with the user experience features that we outlined in Section~\ref{Sec:SocialAcceptance}. Similarly, Table~\ref{Tab:BasicAttributes} outlines attributes that we do not consider as direct determinants of social acceptance but are theorized to significantly impact the representative ones. We refer to these attributes as basic attributes. Notably, a representative attribute can also influence another representative attribute; in such cases, it is considered a basic attribute within that specific relationship.

\begin{table}[htbp]
\caption{Representative Attributes of Social Acceptance}
\label{Tab:RepresentativeAttributes}
\resizebox{\columnwidth}{!}{%
\begin{tabular}{|l|l|l|}
\hline
\textbf{Label} &  \textbf{Attribute} & \textbf{Definition}
\\ \hline
DEP & Dependability & \begin{tabular}[t]{@{}l@{}} Users’ confidence in the agent’s ability to consistently deliver its intended level \\of service \end{tabular} 
\\ \hline
ITU & \begin{tabular}[t]{@{}l@{}} Intention to \\Use \end{tabular} & \begin{tabular}[t]{@{}l@{}} Users' intention to use the agent over a longer period in time \end{tabular} 
\\ \hline
PE & \begin{tabular}[t]{@{}l@{}} Perceived \\Enjoyment \end{tabular} & \begin{tabular}[t]{@{}l@{}} Feelings of joy/pleasure related to the agent's use \end{tabular} 
\\ \hline
PEOU & \begin{tabular}[t]{@{}l@{}} Perceived \\Ease of Use \end{tabular} & \begin{tabular}[t]{@{}l@{}} The degree to which users believe that using the agent would be free of effort \end{tabular} 
\\ \hline
PU & \begin{tabular}[t]{@{}l@{}} Perceived \\Usefulness \end{tabular} & \begin{tabular}[t]{@{}l@{}} The degree to which users believe that the agent would be useful \end{tabular} 
\\ \hline
SEC & Security & \begin{tabular}[t]{@{}l@{}} Perceived protection of the agent and users' data from unauthorized access \\and threats \end{tabular} 
\\ \hline
SI & \begin{tabular}[t]{@{}l@{}} Social \\Influence \end{tabular} & \begin{tabular}[t]{@{}l@{}} Users' perception that other people think they should (not) use the agent \end{tabular} 
\\ \hline
\end{tabular}%
}
\end{table}

\begin{table}[htbp]
\caption{Basic Attributes}
\label{Tab:BasicAttributes}
\resizebox{\columnwidth}{!}{%
\begin{tabular}{|l|l|l|}
\hline
\textbf{Label} & \textbf{Attribute} & \textbf{Definition}
\\ \hline
ANX & Anxiety & \begin{tabular}[t]{@{}l@{}} Evoking anxious or emotional reactions while using the agent \end{tabular} 
\\ \hline
ATT & \begin{tabular}[t]{@{}l@{}} Attitude \\towards \\Technology \end{tabular} & \begin{tabular}[t]{@{}l@{}} Positive or negative feelings about the appliance of the technology \end{tabular} 
\\ \hline
FC & \begin{tabular}[t]{@{}l@{}} Facilitating \\conditions \end{tabular} & \begin{tabular}[t]{@{}l@{}} Factors in the environment that facilitate agent's use \end{tabular} 
\\ \hline
IP & \begin{tabular}[t]{@{}l@{}} Informational \\Privacy \end{tabular} & \begin{tabular}[t]{@{}l@{}} Users' perception of the agent's potential to expose/misuse their personal data to \\unauthorized/irrelevant third parties \end{tabular} 
\\ \hline
PA & \begin{tabular}[t]{@{}l@{}} Perceived \\Adaptiveness \end{tabular} & \begin{tabular}[t]{@{}l@{}} The perceived ability of the agent to adapt to the users' needs \end{tabular} 
\\ \hline
PS & \begin{tabular}[t]{@{}l@{}} Perceived \\Sociability \end{tabular} & \begin{tabular}[t]{@{}l@{}} The perceived ability of the agent to perform sociable behavior \end{tabular} 
\\ \hline
SP & \begin{tabular}[t]{@{}l@{}} Social \\Presence \end{tabular} & \begin{tabular}[t]{@{}l@{}} The experience of sensing a social entity when interacting with the agent \end{tabular} 
\\ \hline
TRU & Trust & \begin{tabular}[t]{@{}l@{}}The belief that the agent performs with personal integrity and reliability \end{tabular} 
\\ \hline
\end{tabular}%
}
\end{table}

\begin{figure*}[ht!]
\centering
\includegraphics[width=9.5cm,keepaspectratio]{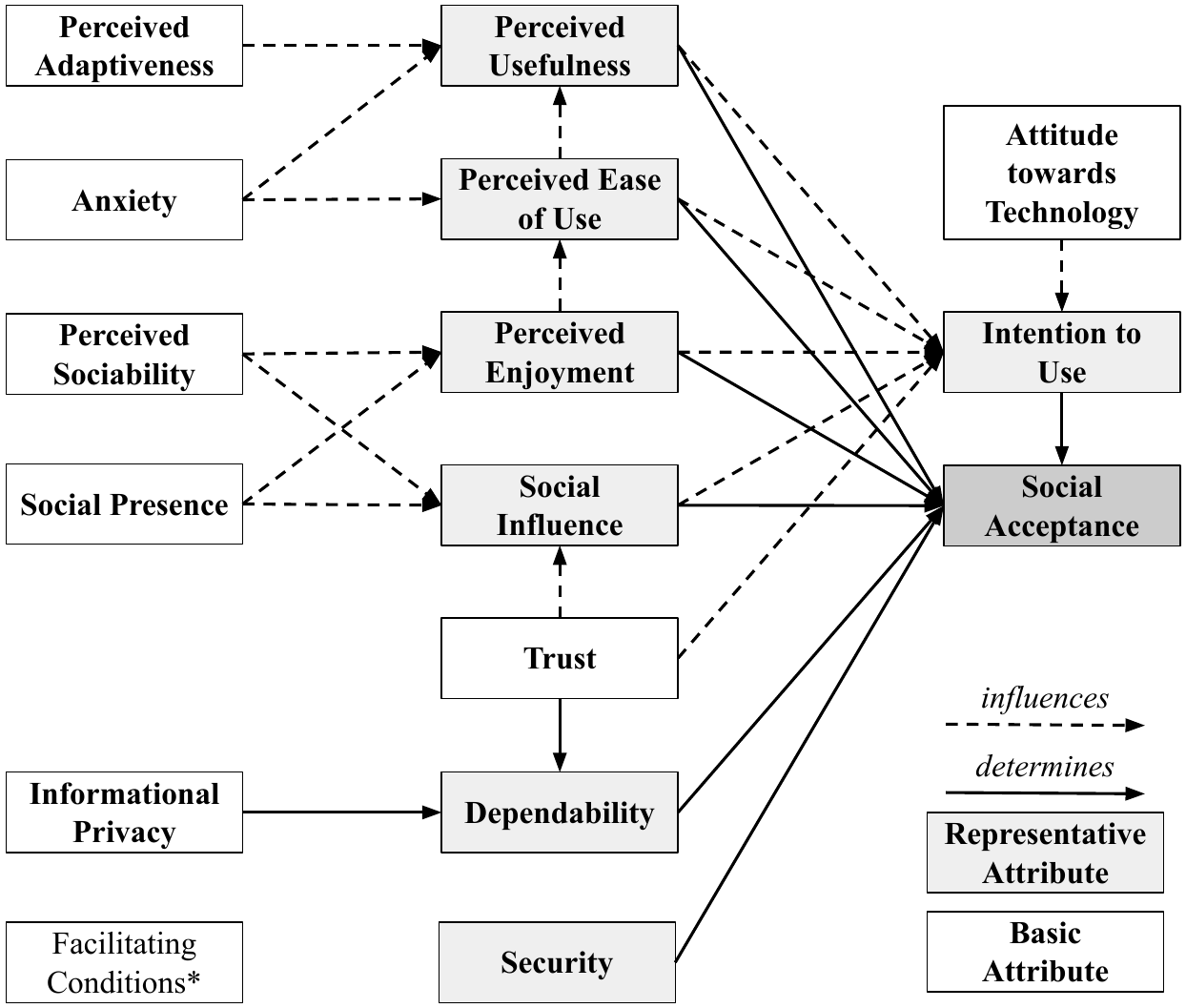}
\caption{Interrelations between the Attributes.} 
\label{Fig:SocialAcceptance}
\end{figure*}

Figure~\ref{Fig:SocialAcceptance} illustrates the model, highlighting the following relationships between the aforementioned attributes, which we either adopted or adapted from the Almere model and analyzed in our experiments (We distinguish between the terms \emph{determined} and \emph{influenced} based on the roles of the attributes. Specifically, attribute A is \emph{determined} by attribute B means that A does not have an independent existence in the model; rather, its value is entirely derived from B. In contrast, attribute A is \emph{influenced} by attribute B means that A maintains its significance in the model, while still being affected by B):
\begin{enumerate}[noitemsep]
    \item Social Acceptance is determined by \emph{(a)} Dependability, \emph{(b)}~Intention to Use, \emph{(c)} Perceived Ease of Use, \emph{(d)} Perceived Enjoyment, \emph{(e)} Perceived Usefulness, \emph{(f)} Security, and \emph{(g)} Social Influence.
    \item Intention to Use is influenced by \emph{(a)} Attitude towards Technology, \emph{(b)} Perceived Ease of Use, \emph{(c)} Perceived Enjoyment, \emph{(d)} Perceived Usefulness, \emph{(e)} Social Influence, and \emph{(f)} Trust.
    \item Perceived Usefulness is influenced by \emph{(a)} Anxiety, \emph{(b)} Perceived Adaptiveness, and \emph{(c)}~Perceived Ease of Use.
    \item Perceived Ease of Use is influenced by \emph{(a)} Anxiety and \emph{(b)}~Perceived Enjoyment.
    \item Perceived Enjoyment is influenced by \emph{(a)} Perceived Sociability and \emph{(b)} Social Presence.
    \item Social Influence is influenced by \emph{(a)} Perceived Sociability, \emph{(b)}~Social Presence, and \emph{(c)}~Trust.
    \item Dependability is determined by \emph{(a)} Informational Privacy and \emph{(b)} Trust.
\end{enumerate}

Facilitating Conditions were not included in this model, as our experiment did not evaluate the actual voluntary use of the system over an extended period, which we aim to evaluate in the future. Instead, the system was used only during a controlled session involving a one-time interaction. Without data on user behavior over time in a naturalistic setting, it is not possible to assess how Facilitating Conditions affect social acceptance or other attributes. Although the Almere model provides a basis for our work, we adapted it to fit our specific study context, evaluating social acceptance of XR agent technology. In contrast to the socially assistive setting of the original model, certain attributes played different roles or interacted differently. For instance, Attitude toward Technology was not linked to Anxiety, Perceived Adaptivity, or Social Influence. Given the professional training context, Social Influence was not examined as a factor shaping user attitudes, and Anxiety was minimal, as participants were generally familiar with digital tools.

Tables~\ref{Tab:QuantitativeQuestionnaire1} and ~\ref{Tab:QuantitativeQuestionnaire2} show the questionnaire we considered based on the Almere model, with some questions adapted to our specific case study and additional questions included to cover other relevant attributes like Security, Informational Privacy, and Trust. Table~\ref{Tab:QualitativeQuestionnaire} presents the questions used for the qualitative interviews.

\begin{table}[ht!]
\caption{Questionnaire for Quantitative Studies}
\label{Tab:QuantitativeQuestionnaire1}
\resizebox{\columnwidth}{!}{%
\begin{tabular}{|l|l|}
\hline
\textbf{Label} & \textbf{Question}
\\ \hline
\multirow[t]{3}{*}{ANX} & \begin{tabular}[t]{@{}l@{}}1. If I should use the virtual trainer, I would be afraid to make mistakes with it.\end{tabular}
\\ 
& 2. I find the virtual trainer scary.
\\
& 3. I find the virtual trainer intimidating.
\\ \hline
\multirow[t]{3}{*}{ATT} & 4. I think it’s a good idea to use the virtual trainer. 
\\
& 5. The virtual trainer would make learning more interesting. 
\\
& 6. It’s good to make use of the virtual trainer.
\\ \hline
\multirow[t]{2}{*}{FC} & 7. I have everything I need to use the virtual trainer. 
\\  
& 8. I know enough of the virtual trainer to make good use of it.
\\ \hline
\multirow[t]{5}{*}{IP} & \begin{tabular}[t]{@{}l@{}}9. I think the virtual trainer could over-collect/misuse my personal information.\end{tabular} 
\\ 
& \begin{tabular}[c]{@{}l@{}}10. It is highly probable that I will share my personal information with this virtual trainer.\end{tabular} 
\\  
& \begin{tabular}[c]{@{}l@{}}11. I am concerned about the potential exposure of sensitive data through this virtual trainer to \\unauthorized parties.\end{tabular} 
\\ 
& \begin{tabular}[c]{@{}l@{}}12. I am concerned that the virtual trainer leaks my personal information to irrelevant third \\parties.\end{tabular} 
\\ 
& \begin{tabular}[c]{@{}l@{}}13. I am concerned that utilizing this virtual trainer could result in misuse of my personal \\information.\end{tabular}
\\ \hline
\multirow[t]{3}{*}{ITU} & \begin{tabular}[c]{@{}l@{}}14. I think I’ll use the virtual assistant during a training activity.\end{tabular} 
\\ 
& \begin{tabular}[c]{@{}l@{}}15. I’m certain to use the virtual assistant during training activities.\end{tabular} 
\\ \cline{2-2} 
& 16. I plan to use the virtual assistant during training activities.
\\ \hline
\multirow[t]{3}{*}{PA} & 17. I think the virtual trainer can be adaptive to what I need. 
\\  
& \begin{tabular}[c]{@{}l@{}} 18. I think the virtual trainer will only do what I need at that particular moment.\end{tabular} 
\\ 
& \begin{tabular}[c]{@{}l@{}}19. I think the virtual trainer will help me when I consider it to be necessary.\end{tabular}
\\ \hline
\multirow[t]{5}{*}{PE} & 20. I enjoy the virtual trainer talking to me. 
\\  
& 21. I enjoy doing things with the virtual trainer. 
\\ 
& 22. I find the virtual trainer enjoyable. 
\\  
& 23. I find the virtual trainer fascinating. 
\\  
& 24. I find the virtual trainer boring.
\\ \hline
\multirow[t]{5}{*}{PEOU} & 25. I think I will know quickly how to use the virtual trainer. 
\\ 
& 26. I find the virtual trainer easy to use. 
\\ 
& 27. I think I can use the virtual trainer without any help. 
\\ 
& \begin{tabular}[c]{@{}l@{}}28. I think I can use the virtual trainer when there is someone around to help me.\end{tabular} 
\\ 
& \begin{tabular}[c]{@{}l@{}}29. I think I can use the virtual trainer when I have a good \\manual.\end{tabular} 
\\ \hline
\multirow[t]{4}{*}{PS} & \begin{tabular}[c]{@{}l@{}}30. I consider the virtual trainer pleasant to have a conversation with.\end{tabular}  
\\ 
& 31. I find the virtual trainer pleasant to interact with. 
\\ 
& 32. I feel the virtual trainer understands me. 
\\ 
& \begin{tabular}[c]{@{}l@{}}33. I think the virtual trainer is nice.\end{tabular}
\\ \hline
\multirow[t]{3}{*}{PU} & \begin{tabular}[c]{@{}l@{}}34. I think the virtual trainer is useful to me during training activities.\end{tabular} 
\\ 
& \begin{tabular}[c]{@{}l@{}}35. It would be convenient for me to have the virtual trainer during training activities.\end{tabular} 
\\ 
& \begin{tabular}[c]{@{}l@{}}36. I think the virtual trainer can help me with different activities.\end{tabular}
\\ \hline
SEC & \begin{tabular}[t]{@{}l@{}}37. I am concerned about the potential for unauthorized users to access this virtual trainer.\end{tabular}
\\ \hline
\multirow[t]{2}{*}{SI} & \begin{tabular}[t]{@{}l@{}}38. I think the training facilitators would like me using the virtual trainer.\end{tabular}
\\ 
& \begin{tabular}[c]{@{}l@{}}39. I think it would give a good impression if I use the virtual trainer.\end{tabular}
\\ \hline
\multirow[t]{5}{*}{SP} & \begin{tabular}[t]{@{}l@{}}40. When interacting with the virtual trainer I felt like I’m talking to a real person.\end{tabular} 
\\ 
& \begin{tabular}[t]{@{}l@{}}41. It sometimes felt as if the virtual trainer was really looking at me.\end{tabular}  
\\ 
& 42. I can imagine the virtual trainer to be a living creature. 
\\ 
& \begin{tabular}[c]{@{}l@{}}43. I often think the virtual trainer is not a real person.\end{tabular} 
\\ 
& 44. Sometimes the virtual trainer seems to have real feelings.
\\ \hline
\end{tabular}%
}
\end{table}

\begin{table}[ht!]
\caption{Questionnaire for Quantitative Studies (Continued from Table~\ref{Tab:QuantitativeQuestionnaire1})}
\label{Tab:QuantitativeQuestionnaire2}
\resizebox{\columnwidth}{!}{%
\begin{tabular}{|l|l|}
\hline
\textbf{Label} & \textbf{Question}
\\ \hline
\multirow[t]{7}{*}{TRU} & 45. I would trust the virtual trainer if it gave me advice. 
\\ 
& \begin{tabular}[c]{@{}l@{}}46. I would follow the advice and trust the information the virtual trainer gave me.\end{tabular} 
\\ 
& \begin{tabular}[c]{@{}l@{}}47. I believe that only legitimate individuals can access this virtual trainer.\end{tabular} 
\\ 
& \begin{tabular}[c]{@{}l@{}}48. I am sure that this virtual trainer is maintaining a secure interaction environment.\end{tabular} 
\\ 
& \begin{tabular}[c]{@{}l@{}}49. I am confident that my anonymity is protected by this virtual trainer.\end{tabular}  
\\ 
& 50. I believe that the virtual trainer acted in my best interest. 
\\ 
& \begin{tabular}[c]{@{}l@{}}51. This virtual assistant/post office agent performed its role of offering services (e.g., teaching about \\a necessary preparation step) really well.\end{tabular}
\\ \hline
\end{tabular}%
}
\end{table}

\begin{table}[htbp]
\caption{Questionnaire for Qualitative Interviews}
\label{Tab:QualitativeQuestionnaire}
\resizebox{\columnwidth}{!}{%
\begin{tabular}{|l|l|}
\hline
\textbf{Label} & \textbf{Question}
\\ \hline
Q01 & How would you describe your experience with the training?
\\ \hline
Q02 & \begin{tabular}[t]{@{}l@{}}Please use 3 words (preferably adjectives) to describe your experience.\end{tabular}
\\ \hline
Q03 & What did you like the most about the training? Why?
\\ \hline
Q04 & What did you like the least about the training? Why?
\\ \hline
Q05 & \begin{tabular}[t]{@{}l@{}}Why did you mostly interact via voice or text with \emph{Guardia}? (Depending on participant's behaviour)\end{tabular}
\\ \hline
Q06 & \begin{tabular}[t]{@{}l@{}}Which specific improvements could make the training better? (Ask specifically for the avatar if it is \\not mentioned by the participants)\end{tabular}
\\ \hline
Q07 & \begin{tabular}[t]{@{}l@{}}How did you feel about the security of the virtual training environment? For example, how confident \\are you that the training content cannot be tampered with?\end{tabular}
\\ \hline
Q08 & \begin{tabular}[t]{@{}l@{}}Do you have any concerns about where and how personal data is stored after the training sessions? \\If the answer is YES, then please specify.\end{tabular}
\\ \hline
Q09 & \begin{tabular}[t]{@{}l@{}}How would you rate the trustworthiness of the avatar compared to training tools/methods you have \\used before?\end{tabular}
\\ \hline
Q10 & \begin{tabular}[t]{@{}l@{}}Anything last to add? (A last question specific to the participant)\end{tabular}
\\ \hline
\end{tabular}%
}
\end{table}

\subsection{Data Analysis}
\label{Subsec:DataAnalysis}

To analyze responses from the quantitative questionnaire, we calculated weighted mean scores (also referred to as adjusted mean scores) for each representative attribute. This was done by aggregating the mean scores of the representative attributes (i.e., non-adjusted mean scores) and their corresponding influencing basic attributes, based on the responses from 25 participants. Mean scores provide a precise measure of central tendency, while weighted mean scores account for the relative importance of each component. This offers a rigorous evaluation by reflecting both the response distribution for each representative attribute and the varying influence of the basic attributes in the final evaluation.

To compute the final, adjusted mean score $V_R$ for each representative attribute, we used the following equation:
\vspace{-0.5mm}
\begin{equation}
    V_R = A_R \cdot w_R + \sum_{p=1}^{x} \left( A_{B_p} \cdot \frac{E_p}{\sum E} \right) \cdot w_B
    \label{Eq:WeightedValue}
\end{equation}
\vspace{-0.5mm}
where
\begin{itemize}[noitemsep]
    \item $A_R$ and $A_{B_p}$ respectively denote the mean scores for the representative and $p$th basic attribute;
    \item $E_p$ denotes the effect size of the basic attribute $p$, and $E$ is the list of effect sizes of all basic attributes; and
    \item $w_R$ and $w_B$ respectively denote the weights of the representative and basic attributes, where $w_R$ + $w_B$ = 1.
\end{itemize}

To compute the mean score of the representative attribute $A_R$, we used $R$ as the list of scores recorded for the representative attribute and $m$ as the number of recorded scores, as follows: 
\vspace{-0.5mm}
\begin{equation}
    A_R = \frac{1}{m} \sum_{i=1}^{m} R_i
    \label{Eq:MeanRepresentative}
\end{equation}

Similarly, for each basic attribute $p$ (out of $x$ basic attributes, which influence the representative attribute's outcome) with $n$ recorded scores, we calculated the average score $A_{B_p}$ of the basic attribute as follows:
\begin{equation}
    A_{B_p} = \frac{1}{n} \sum_{j=1}^{n} B_{p_j}
    \label{Eq:MeanBasic}
\end{equation}

Using pre-defined weights $w_R$ = 0.6 and $w_B$ = 0.4, which we chose based on the relative importance of representative and basic attributes, we determined the final mean score $V$ by incorporating the effect sizes of basic attributes using the Equation~\ref{Eq:WeightedValue}. Notably, in the case of Security as a representative attribute, since there are no basic attributes determining or influencing it, we set $w_R$ = 1 and $w_B$ = 0. Similarly, for Dependability as a representative attribute, since it has no independent measurement and is derived from Informational Privacy and Trust, we set $w_R$ = 0 and $w_B$ = 1. The effect sizes $E_p$ are adopted from the final Almere model. For any new inter-relations introduced in this work, we assumed a uniform distribution for the effect sizes of the basic attributes.

\section{Findings}
\label{Sec:Findings}

\begin{figure*}[ht!]
\centering
\includegraphics[width=9cm,keepaspectratio]{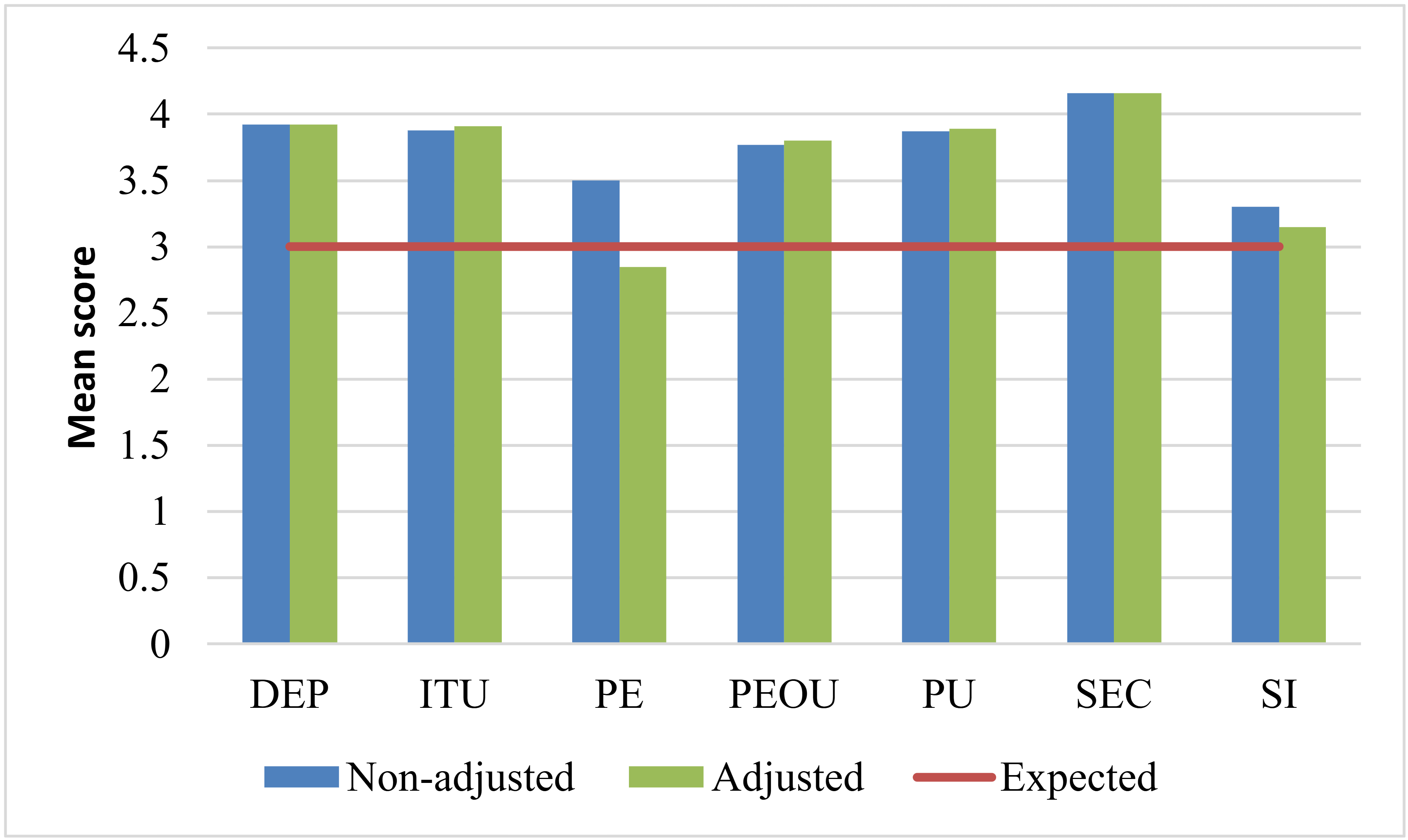}
\caption{`Non-adjusted' and `Adjusted' Mean Scores for the Representative Attributes of Social Acceptance.} 
\label{Fig:MeanScore}
\end{figure*}

Table~\ref{Tab:MeanScoreStandardDeviation} presents the mean scores and standard deviations for each attribute included in the study, based on responses from 25 participants. Each row corresponds to a specific attribute label, where the mean score reflects participants' overall perception, and the standard deviation indicates the variability of responses for that attribute. Upon analyzing the results of the user studies, we found that most of the basic attributes (cf. Table~\ref{Tab:BasicAttributes}) have a mean score above the neutral value of 3.0, showing that the participants had a high positive perception toward these attributes, except for Social Presence (mean score 1.66). This indicates that most participants did not perceive the virtual agent \emph{Guardia} as particularly human-like. Since Social Presence is factored into the non-adjusted mean score for Perceived Enjoyment, the low mean score of the former causes the adjusted mean score for the latter to fall from 3.50 to 2.85 (see Figure~\ref{Fig:MeanScore}). Another factor influencing Perceived Enjoyment is Perceived Sociability (mean score 3.29) (see Table~\ref{Tab:Results}); however, Social Presence contributes to a negative impact overall. This is due to the lower mean score of Social Presence compared to Perceived Sociability, despite both attributes having the same weight. The overall decline and participants' responses in the qualitative interview indeed suggest an opportunity for improvement; for instance, by enhancing the virtual avatar's human likeness, particularly in body language and dialogue. By incorporating more natural, expressive gestures and adding elements of personal engagement, such as small talk, the interaction could feel more authentic and engaging, helping to match users' expectations and improve overall acceptance. 

Notably, we observed an inconsistency in the responses of 3 participants for a pair of opposite questions for Social Presence (see questions 40 and 43 in Table~\ref{Tab:QuantitativeQuestionnaire2}). This is attributed to a lack of engagement while responding to the questionnaire and to misinterpretation of the questions. As noted in Section~\ref{Subsec:MethodologyAndProcedure}, such questions were included to help validate the coherence of participants' responses. To address the issue of misinterpretation, we aim to simplify the language of the questions in future iterations of the user study. 

We also observed that 7 out of 25 participants in the case of Perceived Ease of Use and 8 out of 25 in the case of Perceived Adaptiveness gave identical answers to complementary questions. This suggests that these attributes should be interpreted with a bit more scrutiny. However, since these attributes' mean scores are based on a smaller set of questions, i.e., 5 and 3 questions respectively, the results can still be considered reliable. This is further supported by our observation that even when adjusting the conflicting responses to reflect a more negative evaluation (e.g., assigning a score of 1 to a positive question when a participant gave a score of 5 to its negative counterpart), the mean score only dropped from 3.77 to 3.59 for Perceived Ease of Use, and from 3.44 to 3.20 for Perceived Adaptiveness, while both still remain above 3.0.

\begin{table}[htbp]
\centering
\caption{Attributes' Non-adjusted Mean Score and Standard Deviation}
\label{Tab:MeanScoreStandardDeviation}
\small{%
\begin{tabular}{|l|l|l|}
\hline
\textbf{Attribute Label} & \textbf{Mean Score} & \textbf{Standard Deviation}
\\ \hline
ANX & 4.60 & 0.45
\\ \hline
ATT & 3.92 & 0.88
\\ \hline
FC & 4.10 & 0.71
\\ \hline
IP & 3.98 & 0.72
\\ \hline
ITU & 3.88 & 0.92
\\ \hline
PA & 3.44 & 0.58
\\ \hline
PE & 3.50 & 0.80
\\ \hline
PEOU & 3.77	& 0.50
\\ \hline
PS & 3.29 & 1.00
\\ \hline
PU & 3.87 & 0.91
\\ \hline
SEC & 4.16 & 1.03
\\ \hline
SI & 3.30 & 0.60
\\ \hline
SP & 1.66 & 0.59
\\ \hline
TRU & 3.86 & 0.39
\\ \hline
\end{tabular}%
}
\end{table}

Likewise, the adjusted mean score for Intention to Use and Social Influence showed a minor decline of 0.02 and 0.15, respectively from their non-adjusted mean scores. Here, Social Influence captures the users' perception of the agent concerning social appropriateness, and it is also an influencing factor for the Intention to Use, among others.  

For Dependability, the attributes deriving it are Informational Privacy and Trust. These relate to the collection, handling, and storage of user-specific information, as well as users' confidence in the information provided by the agent \emph{Guardia} during the training sessions. Participants expressed varying levels of concern about privacy and data storage---most of them had none, while some raised concerns related to potential misuse, deployment context, or the sensitivity of the information. A few specifically mentioned risks, such as voice cloning or misuse of personal details. Opinions on trust in the avatar compared to other training tools (e.g., text documents) were mixed. Many participants viewed it as equivalent or superior, citing strengths like a pleasant voice, appropriate tone, and lack of hallucinations or content overload. Others felt supplementary sources were needed. For some participants, trust also depended on factors such as reduced latency and a clearer definition of the training scope. A few participants considered the agent less trustworthy than human trainers, pointing to potential concerns about outdated content and clarity.

\begin{table*}[htbp]
\centering
\caption{`Adjusted' Mean Scores for the Representative Attributes}
\label{Tab:Results}
\resizebox{\columnwidth}{!}{%
\begin{tabular}{|l|l|l|l|l|l|l|l|l|l|}
\hline
\textbf{\begin{tabular}[c]{@{}l@{}}Rep. \\Attribute \\Label\end{tabular}} & \textbf{\begin{tabular}[c]{@{}l@{}}Rep. \\Attribute \\Mean Score \\($A_R$)\end{tabular}} & \textbf{\begin{tabular}[c]{@{}l@{}}Basic \\Attribute \\Label\end{tabular}} & \textbf{\begin{tabular}[c]{@{}l@{}}Basic \\Attribute \\Mean Score \\($A_{B_p}$)\end{tabular}} & \textbf{\begin{tabular}[c]{@{}l@{}}Influencing \\Factor \\($E_p$)\end{tabular}} & \textbf{\begin{tabular}[c]{@{}l@{}}Influencing \\Sum \\($\sum E$)\end{tabular}} & \textbf{\begin{tabular}[c]{@{}l@{}}Basic \\Attribute \\Adjusted \\Mean Score \\(\( A_{B_p} \cdot \frac{E_p}{\sum E} \))\end{tabular}} & \textbf{\begin{tabular}[c]{@{}l@{}}Rep. \\Attribute \\Weight \\($w_R$)\end{tabular}} & \textbf{\begin{tabular}[c]{@{}l@{}}Basic \\Attribute \\Weight \\($w_B$)\end{tabular}} & \textbf{\begin{tabular}[c]{@{}l@{}}Rep. \\Attribute \\Adjusted \\ Mean Score \\($V_R$)\end{tabular}}
\\ \hline
\multirow[t]{2}{*}{DEP} & - & IP & 3.98 & 0.5 & 1 & 1.99 & 0 & 1 & 3.92 
\\ 
& & TRU & 3.86 & 0.5 & & 1.93 & & &
\\ \hline
\multirow[t]{6}{*}{ITU} & 3.88 & ATT & 3.92 & 0.49 & 0.89 & 1.40 & 0.6 & 0.4 & 3.91 
\\ 
& & PE & 3.50 & -0.13 & & 0.32 & & &
\\ 
& & PEOU & 3.77 & -0.11 & & 0.31 & & &
\\ 
& & PU & 3.87 & 0.62 & & 1.74 & & &
\\ 
& & SI & 3.30 & 0.02 & & 0.04 & & &
\\ 
& & TRU & 3.86 & 0.01 & & 0.03 & & &
\\ \hline
\multirow[t]{2}{*}{PE} & 3.50 & PS & 3.29 & 0.10 & 0.79 & 0.41 & 0.6 & 0.4 & 2.85
\\ 
& & SP & 1.66 & 0.69 & & 1.46 & & &  
\\ \hline
\multirow[t]{2}{*}{PEOU} & 3.77 & ANX & 4.60 & 0.27 & 0.84 & 1.47 & 0.6 & 0.4 & 3.80 
\\ 
& & PE & 3.50 & 0.57 & & 2.39 & & &
\\ \hline
\multirow[t]{3}{*}{PU} & 3.87 & ANX & 4.60 & 0.33 & 1.08 & 1.39 & 0.6 & 0.4 & 3.89
\\ 
& & PA & 3.44 & 0.34 & & 1.08 & & & 
\\ 
& & PEOU & 3.77 & 0.41 & & 1.45 & & &
\\ \hline
SEC & 4.16 & - & - & - & - & - & 1 & 0 & 4.16
\\ \hline
\multirow[t]{3}{*}{SI} & 3.30 & PS & 3.29 & 0.33 & 0.99 & 1.10 & 0.6 & 0.4 & 3.15 
\\ 
& & SP & 1.66 & 0.33 & & 0.55 & & &
\\ 
& & TRU & 3.86 & 0.33 & & 1.29 & & & 
\\ \hline
\end{tabular}%
}
\end{table*}

Security was associated with the potential risk of unauthorized access to user data, a concern that was addressed during the system's development. Participants generally responded positively to the implemented security features, as reflected in both the quantitative questionnaire and qualitative interviews. When asked about the security of the virtual training environment, 20 participants considered it conditionally secure, likely due to concerns about the use of an external server or their level of trust in the media organization. Three participants either assumed the platform was secure or had not considered security, due to general trust in the system without awareness of the underlying technical infrastructure. Two participants expressed mixed opinions, indicating uncertainty or a lack of clear information about the system’s security measures.

Nevertheless, participants had an overall pleasant and comprehensible experience during their interactions with the agent, \emph{Guardia}, describing the virtual avatar-based security training as informative, entertaining, and interesting in their responses to the open-ended questions. In summary, of the 25 participants, 44\% reported a positive experience, 8\% reported a negative experience, and 48\% expressed mixed feelings. Participants particularly appreciated Guardia's conversational ability, noting that the chatbot could provide relevant answers beyond the lesson content, which made the experience feel intelligent and personalized. Many valued the sense of guidance and support, expressing that it felt like someone was accompanying them through the training. The combination of voice and text was highlighted as especially helpful for retention and engagement, with several participants praising the pleasant and appealing voice as well as the personal tone of the content. Additionally, users appreciated the lack of fatigue compared to human trainers, and found the avatar’s personality and interactivity engaging, though one participant mentioned that the arms appeared robotic. From a functionality point of view, participants found the User Interface (UI) easy to use; however, they also provided some suggestions for its improvement, for instance, the relative positioning of the avatar and the textbox, and the inclusion of illustrations and videos.

These findings highlight the importance of analyzing the mean scores of the basic attributes, as understanding their influence on the representative attributes of social acceptance can help identify areas for system improvement. 

\section{Conclusion and Future Work}
\label{Sec:Conclusion&FutureWork}

We presented the results of a user study that we conducted to evaluate the users' social acceptance of XR agent technology---specifically, a virtual conversational agent used in the context of journalism training. To this end, we adapted and extended the Almere model to design a questionnaire focusing on seven key attributes: Dependability, Intention to Use, Perceived Enjoyment, Perceived Ease of Use, Perceived Usefulness, Security, and Social Influence. To complement the quantitative data, we also conducted interviews to gather more detailed, nuanced responses from participants. The findings contribute to the understanding of how users perceive such a solution, while also highlighting opportunities for further development, in particular, in improving the avatar’s behavior to appear more human-like and relatable in the considered social context. Notably, we did not include the responses to the questions about Facilitating Conditions in our analysis. In future work, we aim to explore the potential impact of this construct on social acceptance and/or other attributes affecting social acceptance.

We also plan to conduct a second round of user studies with an improved version of the agent prototype. Based on the findings of the current study, these improvements will primarily focus on enhancing the human-likeness of the avatar, refining the content of the training, and upgrading the user interface to better support seamless interaction between the avatar and human users. These steps are expected to further improve user engagement and acceptance as the system evolves.

We also plan to extend our analysis to other social contexts, such as post offices, where such agents can assist customers. To this end, we have recently conducted two preliminary pilot studies with a postal service provider, which, besides providing postal services, offers communications, postal savings products, logistics, and financial and insurance services, to explore the system's applicability in service-oriented XR scenarios. The first pilot focused on a Customer Reception Kiosk, an XR system deployed at the postal company's headquarters to enhance reception services through intuitive touch-based and voice-based interactions. The second pilot involved a Virtual Assistant for Info Point and Service Offering, designed for dynamic, conversational engagement, including the ability to physically approach users through a mobile, robotic module. Unlike the user study presented in this paper, which was conducted within a media organization and where participants were exclusively employees, the postal company’s pilot studies involved a broader and more socially diverse user base, including individuals with varying levels of experience with virtual agents---from naive to experienced users. We do not discuss the results of these preliminary user studies in this paper as they are not mature enough. We will carry out additional work to integrate and analyze them to assess the generalizability of our findings, which will help us better understand how users in different social contexts interact with the system and further evaluate its acceptance.

\section*{Acknowledgement}

This work was supported by the Horizon Europe program under the Grant Agreement 101070351 (``SERMAS: Socially-acceptable eXtended Reality Models and Systems'') and by Innovate UK.

\bibliographystyle{plain}
\bibliography{references}

\begin{extended}

\appendix



\section*{Supplementary material}

# A  User Consent Form

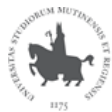

COMITATO ETICO PER LA RICERCA DELL'UNIVERSITÀ DI MODENA E
REGGIO EMILIA (CEAR)
VIA DELL'UNIVERSITÀ' 4
41124 MODENA, ITALIA
CEA.RICERCA@PEC.UNIMORE.IT
CEA.RICERCA@UNIMORE.IT

**INFORMED CONSENT TO PARTICIPATE IN RESEARCH**

TITLE OF THE STUDY:

**Socially-acceptable Extended Reality Models and Systems**

**General Information**

Dear Sir/Madam,

We would like to propose you participate in a research project. You have the right to be informed about the purpose, characteristics, and procedures of the study so that you can take an informed and free decision to participate. Please read the following carefully. The researchers involved in this project are available to answer your questions.

**Head of the study**

| Prof. Lorenzo Sabattini | +39-0522-523528 |
|---|---|
| (name) | (phone number) |

**Researcher**

| Axel Primavesi | +49172 2526049 |
|---|---|
| (name) | (phone number) |

**What is the purpose of this study?**
The general aim of this study is to test the three use cases proposed in the European project SERMAS, namely the POST OFFICE AGENT, the RECEPTIONIST AGENT, and the SECURITY AGENT, and to assess the social acceptance of these technologies by users.

**How will the study be conducted?**
The study will be conducted in one of the following locations: at the Tecnopolo in Reggio Emilia and in the headquarters of two project partners, Poste Italiane (Rome) and Deutshe Welle (Bonn and Berlin). Different technologies will be tested at different sites.

Specifically, at Poste Italiane, both the POST OFFICE and the RECEPTIONIST agents will be tested using two different interaction scenarios. In the first scenario, the POST OFFICE AGENT will be used in different interaction tasks, such as question/answer, operational tasks (e.g., sending a package, issuing tickets and booking consultations), informative tasks (automatic compilation of the ISEE request, info on products and services, support for form filling) and orientation within the office (map view).

1

In the second scenario, the RECEPTIONIST AGENT will be used in activities that involve the reception of guests who have an appointment with Poste Italiane staff or employees. In particular, the scenario includes reception and control procedures to ensure guests' access to the building. The participant will be asked to interact with the agent to carry out the check-in procedures and be accompanied by the agent at the meeting point

At the Deutsch Welle headquarters, a SECURITY AGENT will be tested. The agent will act as a virtual trainer for journalists on three specific topics:
1. The behaviour to be kept at checkpoints.
2. On reporting from political demonstrations.
3. The preparation of a running bag for situations of emergency.

**Who do we propose to participate?**
Only individuals older than 18 years of age can participate in the study.

**Is participation in the study compulsory?**
Your participation is completely voluntary, and you have the right to agree or refuse to participate in this study. Furthermore, if you change your mind and want to withdraw, **you are free to do so at any time and without having to provide any explanation.**

**What are the steps to participate in the study?**
Before participating, you will receive detailed information on the characteristics, risks, and benefits of the study. At the end of the information phase, you can agree to participate in the study by signing the informed consent form. Only after you have given your written consent can you actively participate in the proposed study.

**What will you be asked to do?**
As regards the study taking place in Poste Italiane, users will be asked to interact with the RECEPTIONIST AGENT and/or the POST OFFICE AGENT, by voice or text., to perform some tasks. At the end of the scenario, users will be asked to complete a questionnaire on behavioral and functionals aspects of the agent, and on technology security, trust and privacy.

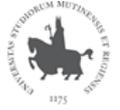

As regards the study taking place in Deutsche Welle headquarters, users will be required to interact with the SECURITY AGENT via voice or text. At the end of the scenario, users will be asked to complete a questionnaire on behavioral and functionals aspects of the agent, and on technology security, trust and privacy.

**What are the possible risks of the study?**
There are no known risks.

**What are the possible benefits to the participants?**
There is little direct benefit from participating in this study. However, the study will provide useful information to guide the design of assistive robots and virtual agents and increase their social acceptance.

**What happens if information about your health emerges during the course of the study?**
If the study reveals information that may be useful for your health, you can choose whether or not to be informed in the "Informed consent" section.

**How is the confidentiality of information guaranteed?**
Before starting the experiment, you will be asked to respond to some demographic questions. This information, along with all data emerging from the study, is important for the research and will be protected. Your identity will not be stored with your data. Instead, your identity will be assigned a code number. Stored data will be completely anonymous.

**How will your personal data be used?**
The data collected will be used in an anonymous and aggregated form, so that it is not possible to trace the data of individuals for thesis work and/or scientific publications, in accordance with what is established in "Consent to the processing of personal data for scientific purposes" which you will sign separately if you decide to participate.

**Other important information**
Please be informed that the study will be conducted in accordance with the ethical principles established by the "World Medical Association. The Helsinki Declaration: Ethical principles for biomedical research involving human beings" (Helsinki Declaration) and in the "Convention on Human Rights and Biomedicine" (Oviedo Convention).

The original of the Informed Consent, signed by you, will be kept by the person responsible for this study, and you are entitled to receive a copy.

**We thank you for your availability and help**

**STATEMENT OF THE RESEARCHER**

I declare that I have provided the participant with full information and detailed explanations about

3

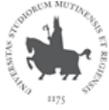

the nature, purpose, procedures, and duration of this research study.

I also declare that I have provided the participant with the information sheet.

AXEL PRIMAVESI

_____

SIGNATURE OF THE RESEARCHER                     Date

_____

**INFORMED CONSENT SIGNATURE**

I declare that I have received information which has enabled me to understand the research project, also considering additional clarifications requested by me. I confirm that I have received a copy of this information document.

SIGNATURE                                       Date

_____

COMITATO ETICO PER LA RICERCA DELL'UNIVERSITÀ DI MODENA E
REGGIO EMILIA (CEAR)
VIA DELL'UNIVERSITÀ' 4
41124 MODENA, ITALIA
CEA.RICERCA@PEC.UNIMORE.IT
CEA.RICERCA@UNIMORE.IT

**INFORMED CONSENT**

Participant's code _____

I,

Name:                           Surname:

hereby declare:

- That I have received a full explanation regarding the request to participate in this study and sufficient information about the risks and benefits involved in the study as set out in the attached information sheet.

- That I have been able to discuss the explanations, to ask all the questions considered necessary and that I received satisfactory answers on this matter.

- That I was also informed of my right to withdraw from the research at any time.

- That the participation in the study is completely voluntary.

- That I have been/and I am informed and I agree to my data being made available not only to the study directors and their delegates, but also to the Ethics Committee, if requested; and I have been/ am also informed that my data may be the subject of disclosure to national and international scientific congresses or publication for scientific reasons in national and international journals, but that in any case my identity will be protected by confidentiality (the data will always be used in an ANONYMOUS and AGGREGATED form).

- That I have been/and I am also informed of my right to have free access to the documentation related to the trial and the evaluation expressed by the Ethics Committee.

- That I have been given a copy of this consent.

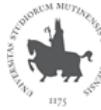

Therefore, in the light of the information provided to me (please select):

| | | | | |
|---|---|---|---|---|
| ☐ | I AGREE | ☐ | I DO NOT AGREE | To participate to the study |
| ☐ | I AGREE | ☐ | I DO NOT AGREE | To audio and/or video* recordings<br><br>*It does not apply to Deutshe Welle scenario |
| ☐ | I WANT | ☐ | I DO NOT WANT | To be informed about any results of the study that are useful to my health. If you wish to be informed, please indicate a telephone contact: |
| ☐ | I AGREE | ☐ | I DO NOT AGREE | To be contacted again |

_____
PLACE AND DATE          SIGNATURE OF THE PARTICIPANT


_____
PLACE AND DATE          SIGNATURE OF THE RESEARCHER

**PRIVACY STATEMENT**

Dear Sir/Madam,

We inform you, pursuant to art. 13 of EU Regulation 2016/679 (hereinafter, GDPR), that the processing of your personal data, information about you, other special categories of data, collected during the study, will be based on respect for fundamental rights and freedoms and the principles of correctness, lawfulness, transparency, minimization of data, accuracy, integrity and confidentiality and may be performed manually or electronically or in any case with the aid of computerized or automated tools.

In particular, with reference to personal data revealing ethnic origin, religious beliefs, philosophical views, political opinions, trade union membership, as well as processing genetic data, biometric data, intended to uniquely identify a person, data relating to health or sexual life or sexual orientation, we inform you that:

• The data provided voluntarily will be used only for study and research purposes and will not be communicated or disseminated.

• The data holder, that is the body that determines how and why the data will be processed, is the University of Modena and Reggio Emilia, located in via dell'Università 4, 41121 Modena, in the person of Rector Carlo Adolfo Porro, Legal Representative, (hereinafter: the Holder). You can contact the Data Controller by writing to the physical address above or sending an e-mail to rettore@unimore.it or a PEC to rettore@pec.unimore.it.

• The Data Protection Officer (hereinafter DPO), that is the data protection officer, to whom you can address for all questions related to the exercise of your rights arising from the GDPR is the Avv. Vittorio Colomba; who can be contacted at dpo@unimore.it.

• The data Controller is Prof. Lorenzo Sabattini, Principal Investigator (PI) of the project.

• The provision of data is optional and any refusal to provide such data may only result in your participation in the study/research project being terminated.

• Except for the exceptions provided by the Regulation for the use of data for scientific research purposes (Article 89 of the GDPR and All. A.4 of the Legislative Decree no. 196/2003), you are entitled at any time:

a) to access personal data and obtain confirmation of the existence of personal data

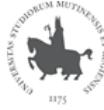

COMITATO ETICO PER LA RICERCA DELL'UNIVERSITÀ DI MODENA E REGGIO EMILIA (CEAR)
VIA DELL'UNIVERSITÀ' 4
41124 MODENA, ITALIA
CEA.RICERCA@PEC.UNIMORE.IT
CEA.RICERCA@UNIMORE.IT

concerning your protected persons;
b) to obtain the rectification or deletion of data or the limitation of their processing;
c) if the data is in electronic form, request portability;
d) object to processing for legitimate reasons;
e) submit a complaint to the supervisory authority.

In this respect, you may assert your rights by contacting the Data Holder and/or the DPO of the University.

• In accordance with the GDPR, data will be stored for a period not exceeding 10 years at the University of Modena and Reggio Emilia under the responsibility of the Data Controller.

Consent to the processing of personal data and particular personal data referred to in art. 9 of the GDPR of adults.

THE UNDERSIGNED
_____

Place of birth _____
Date of birth _____

acquired the information provided by the Data Controller with the information above and aware that the processing will concern "particular personal data referred to in art. 9 of the GDPR",

□ AGREES
□ DOES NOT AGREE

To the processing of data necessary for the study/research project.

Place and date _____
Signature         _____



\end{extended}

\end{document}